\documentclass[12pt]{article}
\newcommand{\newsection}[1]{\section{#1}\setcounter{equation}{0}}

\newcounter{newapp}
\usepackage{amssymb}
\usepackage{euscript}
\usepackage{epsfig,bbm}

\def\a{{\alpha}}
\def\b{{\beta}}
\def\d{\partial}

\def\e{{\epsilon}}
\def\bra#1{\langle #1 |}
\def\ket#1{|#1 \rangle}
\def\0{\nonumber}

\def\det{{\rm det}}
\def\Det{{\rm Det}}
\def\log{{\rm log}}

\def\exp{{\rm exp}}

\newcommand\V{{\cal{V}}}

\newcommand\N{{\cal{N}}}

\newcommand\ee{\end{eqnarray}}      
\newcommand\be{\begin{eqnarray}}
\newcommand\ba{\begin{array}}           
\newcommand\ea{\end{array}}
\newcommand\eeq{\end{equation}}     
\newcommand\beq{\begin{equation}}

\begin{document}
\begin{flushright}
 SISSA/55/04/EP\\ hep-th/0409063
\end{flushright}

\vspace{1.2in}
\begin{center}
{\large\bf Exact time--localized solutions in  Vacuum String Field Theory}
\end{center}
\vspace{0.2in}
\begin{center}
L. Bonora\footnote{ bonora@sissa.it},
C.Maccaferri\footnote{maccafer@sissa.it},
R.J.Scherer Santos\footnote{scherer@sissa.it}, D.D.Tolla\footnote{tolla@sissa.it}\\
\vspace{2mm}

{\it International School for Advanced Studies (SISSA/ISAS)\\
Via Beirut 2--4, 34014 Trieste, Italy, and INFN, Sezione di
Trieste}\\
\end{center}
\vspace{0.3in}
\begin{center}
{\bf Abstract}
\end{center}
We address the problem of finding star algebra projectors
that exhibit localized time profiles. We use the double Wick
rotation method, starting from a Euclidean (unconventional) lump
solution, which is characterized by the Neumann matrix  being the
conventional one for the continuous spectrum, while the inverse of
the conventional one for the discrete spectrum. This is still a
solution of the projector equation and we show that, after inverse
Wick--rotation, its time profile has the desired localized time
dependence. We study it in detail in the low energy regime (field
theory limit) and in the extreme high energy regime (tensionless
limit) and show its similarities with the rolling tachyon
solution.
~\\
\vspace{5cm}
\pagebreak

\newsection{Introduction}

The search for time--dependent solutions has lately become one
of the prominent research topics in string theory. Particularly interesting
is the search for solutions describing the decay of D-branes.
An archetype problem in open bosonic string theory is
describing the evolution from the maximum of the tachyon potential
to the (local) minimum. Such a solution known as rolling tachyon,
if it exists, describes the decay of the space filling
D25--brane corresponding to the unstable perturbative vacuum to the
locally stable vacuum. That such a solution exists has been
argued in many ways, \cite{Senroll}, see also \cite{KMS,LLM,GIR}.
A natural framework where
to study such a nonperturbative problem is String Field Theory
(SFT). But, although there have been some attempts to describe such
phenomena in a SFT framework \cite{Kluson}, no analytical control
has been achieved so far.

In \cite{BMP1} we noticed that in this regard Vacuum String Field Theory
(VSFT) could play an important role.  Let us recall that
VSFT, \cite{Ras}, is a version of Witten's open SFT, \cite{W1}, that is
supposed to describe the  theory at the minimum of the tachyonic potential.
There is evidence that at this point the negative tachyonic potential
exactly compensates  for the D25--brane tension. No open string mode
is expected to be excited, so that the BRST cohomology must be trivial.
This state can only correspond to the closed string vacuum.

Due to these properties VSFT is a simplified
version of SFT. The BRST operator ${\cal Q}$ takes a very simple form
in terms of ghost oscillators alone. It is clearly simpler to work in
such a framework than in the original SFT. In fact many classical
solutions have been shown to exist, which are candidates for representing
D--branes (the sliver, the butterfly, etc.), and other classical
solutions have been found (lump solutions) which may represent
lower dimensional D--branes \cite{KP, RSZ1, RSZ2, MT, FKM, Oka, GRSZ1}.

In the present paper we show that the matter star algebra contains exact time--dependent
projectors with the appropriate characteristic to represent S-branes,
that is solitonic solutions localized in time. We show in fact that
the time profile of such solutions is dominated for large $t$
by a factor $\exp(- at^2)$ with positive constant $a$. At time $t=0$ the solution
takes the form of a deformed sliver (D25-brane), the deformation being
parametrized by two continuous parameters. At infinite future (and infinite
past) time it becomes 0, i.e. it flows into the stable vacuum.
If the initial configuration happens to
coincide exactly with the sliver (no deformation present) there cannot
be any time evolution. Therefore an initial deformation away from the sliver
is essential for true time evolution. Needless to say this is strongly
reminiscent of Sen's rolling tachyon solution, \cite{Senroll} or
of an S--brane, \cite{sbrane}, i.e of a state finely tuned to be poised at
the initial time near the top of the tachyon potential and let free to evolve.

The technique to produce such a solution is based on double Wick--rotation,
as is customary in such kind of trade. Our reference solution is obtained by
picking a Euclidean lump solution with one transverse space direction
(a D24-brane) and then performing an inverse Wick--rotation along such
a direction. However the important ingredient is that our lump solution
is not the ordinary one. Since the spectrum
of the twisted Neumann coefficient matrices of the three strings vertex
nicely split into a continuous and a discrete part, we define a new solution
in which the squeezed state matrix is made of a continuous part, which is the
same as for the conventional lump, and a discrete part which is {\it inverted}
with respect to the ordinary lump. We call this {\it unconventional lump}, see
eqs.(\ref{rollsliv},\ref{T'NM}) below. After inverse--Wick--rotating it
we get the desired time behavior, (\ref{x0yprof}).

In the previous paragraphs we have informally talked about time.
Now we would like to be more precise. Our time is nothing but a
Wick--rotated space coordinate, representing the position of the
string center--of--mass, and it couples to the open string (flat)
metric. In the conclusive section we will discuss a possible
connection of such time with the closed string time (which couples
to the bulk gravity metric). See \cite{MZ,Sen2,GrossE,Ohmori} for
discussions related to the definition of time in SFT.

In this paper we limit ourselves to essentially
one type of solution. It is obvious however that one can describe in
the same way the decay of Dk-branes for any $k<24$. There are also other
types of solutions. They will be illustrated elsewhere
together with complementary aspects of the present solution.

 The paper is organized as follows. In section 2 we review a few basic
 VSFT formulas and briefly introduce the idea of {\it inverse sliver}
 and {\it inverse lump}. In section 3 we review the diagonal representation
 in VSFT and write down some useful approximation formulas. Section 4 is
 of pedagogical nature: we show that if we use the above recipe starting from
 an ordinary lump, we do not get anywhere. This illustrates the need to start
 from unconventional lumps. In section 5 we deal with the main problem,
 that is showing that starting from an unconventional lump we arrive
 at the desired time--dependent solution. In section 6 we describe the low energy and tensionless limits.
 Section 7 is devoted to a discussion of the results.

\newsection{Sliver, inverse sliver and lumps}

In order to render this paper as self--contained as possible, in this section
we collect many well--known formulas and results concerning VSFT.
 The action is
\beq
{\cal S}(\Psi)= - \frac 1{g_0^2} \left(\frac 12 \langle\Psi
|{\cal Q}|\Psi\rangle +
\frac 13 \langle\Psi |\Psi *\Psi\rangle\right)\label{sftaction}
\eeq
where
\beq
{\cal {Q}} =  c_0 + \sum_{n>0} \,(-1)^n \,(c_{2n}+ c_{-2n})\label{calQ}
\eeq

The equation of motion is
\beq
{\cal Q} \Psi = - \Psi * \Psi\label{EOM}
\eeq
and the ansatz for nonperturbative solutions is in the factorized form
\beq
\Psi= \Psi_m \otimes \Psi_g\label{ans}
\eeq
where $\Psi_g$ and $\Psi_m$ depend purely on ghost and matter
degrees of freedom, respectively. Then eq.(\ref{EOM}) splits into
\be
{\cal Q} \Psi_g & = & - \Psi_g *_g \Psi_g\label{EOMg}\\
\Psi_m & = & \Psi_m *_m \Psi_m\label{EOMm}
\ee
where $*_g$ and $*_m$ refers to the star product involving only the ghost
and matter part.\\
The action for this type of solution becomes
\beq
{\cal S}(\Psi)= - \frac 1{6 g_0^2} \langle \Psi_g |{\cal Q}|\Psi_g\rangle
\langle \Psi_m |\Psi_m\rangle \label{actionsliver}
\eeq
$\langle \Psi_m |\Psi_m\rangle$ is the ordinary inner product,
$\langle \Psi_m |$ being the $bpz$ conjugate of $|\Psi_m\rangle$
(see below).

It is well--known how to find solutions to (\ref{EOMg}).
We pick one of them, but since the ghost part will not play a role
we will simply ignore it in the following.
We will instead be concerned with the matter part, eq.(\ref{EOMm}).
The solutions are projectors of the $*_m$ algebra.
The $*_m$ product is defined as follows
\beq
_{123}\!\langle V_3|\Psi_1\rangle_1 |\Psi_2\rangle_2 =_3\!\langle
\Psi_1*_m\Psi_2|
\label{starm}
\eeq
where the three strings vertex $V_3$ is
\beq
|V_3\rangle_{123}= \int d^{26}p_{(1)}d^{26}p_{(2)}d^{26}p_{(3)}
\delta^{26}(p_{(1)}+p_{(2)}+p_{(3)})\,{\rm exp}(-E)\,
|0,p\rangle_{123}\label{V3}
\eeq
with
\beq
E= \sum_{a,b=1}^3\left(\frac 12 \sum_{m,n\geq 1}\eta_{\mu\nu}
a_m^{(a)\mu\dagger}V_{mn}^{ab}
a_n^{(b)\nu\dagger} + \sum_{n\geq 1}\eta_{\mu\nu}p_{(a)}^{\mu}
V_{0n}^{ab}
a_n^{(b)\nu\dagger} +\frac 12 \eta_{\mu\nu}p_{(a)}^{\mu}V_{00}^{ab}
p_{(b)}^\nu\right) \label{E}
\eeq
Summation over the Lorentz indices $\mu,\nu=0,\ldots,25$
is understood and $\eta$ denotes the flat Lorentz metric.
The operators $ a_m^{(a)\mu},a_m^{(a)\mu\dagger}$ denote the non--zero
modes matter oscillators of the $a$--th string, which satisfy
\beq
[a_m^{(a)\mu},a_n^{(b)\nu\dagger}]=
\eta^{\mu\nu}\delta_{mn}\delta^{ab},
\quad\quad m,n\geq 1 \label{CCR}
\eeq
$p_{(a)}$ is the momentum of the $a$--th string and
$|0,p\rangle_{123}\equiv |p_{(1)}\rangle\otimes
|p_{(2)}\rangle\otimes |p_{(3)}\rangle$ is
the tensor product of the Fock vacuum
states relative to the three strings with definite c.m.  momentum .
$|p_{(a)}\rangle$ is
annihilated by
the annihilation
operators $a_m^{(a)\mu}$ ($m\geq1$) and it is eigenstate of the
momentum operator
$\hat p_{(a)}^\mu$
with eigenvalue $p_{(a)}^\mu$. The normalization is
\beq
\langle p_{(a)}|\, p'_{(b)}\rangle = \delta_{ab}\delta^{26}(p+p')\0
\eeq
The symbols $V_{nm}^{ab},V_{0m}^{ab},V_{00}^{ab}$ will denote
the coefficients computed in \cite{CST,GJ1,Ohta, tope, leclair1}.
We will use them in the notation of Appendix A and B of \cite{RSZ2}.

To complete the definition of the $*_m$ product we must specify the
$bpz$ conjugation properties of the oscillators
\beq
bpz(a_n^{(a)\mu}) = (-1)^{n+1} a_{-n}^{(a)\mu}\0
\eeq

Let us now return to eq.(\ref{EOMm}). Its solutions are projectors of
the $*_m$ algebra. The simplest one is the sliver.
In this paper we will discuss a solution describing the decay
of a D25--brane, represented by a sliver. We know that the sliver
solution is in many regards too simple and too singular.
However for our purposes in this paper it is better to avoid
complications such as the addition of a background $B$ field, \cite{BMS}, or
 dressing,  \cite{BMP1,BMP2}. This will allow
us to concentrate on the essential aspects of the problem, avoiding
inessential formal complications. A more thorough treatment will
be given elsewhere.

Let us recall the main points concerning the sliver solution.
It is translationally invariant. As a consequence all momenta
can be set to zero. The integration over the momenta can  be dropped
and the only surviving part in $E$ will be the one involving $V_{nm}^{ab}$.
The sliver is defined by
\beq
|\Xi\rangle = \N e^{-\frac 12 a^\dagger Sa^\dagger}|0\rangle,\quad\quad
a^\dagger S a^\dagger = \sum_{n,m=1}^\infty a_n^{\mu\dagger} S_{nm}
 a_m^{\nu\dagger}\eta_{\mu\nu}\label{Xi}
\eeq
This state satisfies eq.(\ref{EOMm}) provided the matrix $S$ satisfies
the equation
\beq
S= V^{11} +(V^{12},V^{21})(1-
\Sigma{\cal V})^{-1}\Sigma
\left(\matrix{V^{21}\cr V^{12}}\right)\label{SS}
\eeq
where
\beq
\Sigma= \left(\matrix{S&0\cr 0& S}\right),
\quad\quad\quad
{\cal V} = \left(\matrix{V^{11}&V^{12}\cr V^{21}&V^{22}}\right),
\label{SigmaV}
\eeq
The proof of this fact is well--known. First one expresses
eq.(\ref{SigmaV}) in terms of the twisted matrices $X=CV^{11},X_+=CV^{12}$
and $X_-=CV^{21}$, together with $T=CS=SC$, where
$C_{nm}= (-1)^n\delta_{nm}$. The matrices $X,X_+,X_-$ are mutually
commuting. Then, requiring $T$ to commute with them as well, one can show
that eq.(\ref{SigmaV}) reduces to the algebraic equation
\beq
XT^2-(1+X)T+X=0\label{algeq}
\eeq
The sliver solution is
\beq
T= \frac 1{2X} (1+X-\sqrt{(1+3X)(1-X)})\label{sliver}
\eeq

The normalization constant $\N$ is calculated to be
\beq
\N= (\Det (1-\Sigma \V))^{\frac{D}{2}}\label{norm}
\eeq
The contribution of the sliver to the matter part of the action
(see (\ref{actionsliver})) is given by
\beq
\langle \Xi|\Xi\rangle = \frac {\N^2}{(\det (1-S^2))^{\frac{D}{2}}}
\label{ener}
\eeq
Both eq.(\ref{norm}) and (\ref{ener}) are ill--defined and need to be
regularized. As anticipated we will not discuss this point here.

Now let us remark that there is another solution to
(\ref{algeq}), i.e. $1/T$. In fact  (\ref{algeq}) is invariant under
the substitution $T \leftrightarrow 1/T$. $1/T$ is given by
the RHS of eq.(\ref{sliver}) with the -- sign replaced by the + sign in front
of the square root. We will call it the {\it inverse sliver}.
This solution was previously discarded, \cite{RSZ2},
because of the bad asymptotic behaviour of the $1/T$ eigenvalues.
However it is exactly this behaviour that will allow us, in the precise sense
clarified in section 5, to find interesting
time--dependent solutions\footnote{Notwithstanding the divergent behaviour
of the eigenvalues it is perhaps possible to associate a definite meaning
to the energy density of some of these solutions. They may be interesting
as solutions of VSFT also without reference to time dependence.}.

The sliver will be our reference solution in the following, but, in order to build our time-dependent
solution, we will explicitly need another kind of solution, the lump \cite{RSZ2}. The lump solution is
engineered to represent a lower dimensional brane, therefore it will have transverse directions along
which translational invariance is broken. Accordingly we split the three string vertex into the tensor
product of the perpendicular part and the parallel part \be |V_3\rangle = |V_{3,\perp}\rangle \,
\otimes\,|V_{3,_\|}\rangle\label{split} \ee and the exponent $E$, accordingly, as $E=E_\|+ E_\perp$. The
parallel part is the same as in the sliver case while the perpendicular part is modified as follows.
Following \cite{RSZ2}, we denote by $x^\a,p^\a$, $\a=1,...,k$ the coordinates and momenta in the
transverse directions and introduce the zero mode combinations \be a_0^{(r)\alpha} = \frac 12 \sqrt b
\hat p^{(r)\alpha} - i\frac {1}{\sqrt b} \hat x^{(r)\alpha}, \quad\quad a_0^{(r)\alpha\dagger} = \frac
12 \sqrt b \hat p^{(r)\alpha} + i\frac {1}{\sqrt b}\hat x^{(r)\alpha}, \label{osc} \ee where $\hat
p^{(r)\alpha}, \hat x^{(r)\alpha}$ are the zero momentum and position operator of the $r$--th string,
and we have introduced the parameter $b$ as in \cite{RSZ2}. It follows \be
\big[a_0^{(r)\alpha},a_0^{(s)\beta\dagger}\big]= \eta^{\alpha\beta}\delta^{rs} \label{a0a0} \ee Denoting
by $|\Omega_{b}\rangle$ the oscillator vacuum (\,$a_0^\alpha|\Omega_{b}\rangle=0$\,), the relation
between the momentum basis and the oscillator basis is defined by \be |\{p^\alpha\}\rangle_{123} =
\left(\frac b{2\pi}\right)^\frac k4 {\rm exp} \left[\sum_{r=1}^3 \left(- \frac b4 p^{(r)}_\alpha
\eta^{\alpha\beta}p^{(r)}_\beta+ \sqrt b  a_0^{(r)\alpha\dagger}p^{(r)}_\a - \frac 12
a_0^{(r)\alpha\dagger}\eta_{\alpha\beta}a_0^{(r)\beta\dagger} \right)\right]|\Omega_{b}\rangle\0 \ee
Next we insert this equation inside $E'_\perp$ and eliminate the momenta along the perpendicular
directions by integrating them out. The overall result of this operation is that, while
$|V_{3,\|}\rangle$ is the same as in the ordinary case,
\be
|V_{3,\perp}\rangle'= K_k\,
e^{-E'}|\Omega_b\rangle\label{V3'}
\ee
with
\be
&&K_k=\left( \frac {2\pi b^3}{\sqrt{3}(V_{00}+b/2)} \right)^{\frac k4},\label{K2}\\
&&E'= \frac 12 \sum_{r,s=1}^3 \sum_{M,N\geq 0} a_M^{(r)\a\dagger}
V_{MN}^{'rs} a_N^{(s)\b\dagger}\eta_{\a\b}\label{E'}
\ee
where $M,N$ denote the couple of indices $\{0,m\}$ and $\{0,n\}$,
respectively.
The coefficients $V_{MN}^{'rs}$ are given in Appendix B of \cite{RSZ2}.
The new Neumann coefficients matrices $V^{'rs}$ satisfy the same relations as
the $V^{rs}$ ones. In particular one can introduce the matrices $X^{'rs}=
C V^{'rs}$, where $C_{NM}=(-1)^N\, \delta_{NM}$, which turn out to commute with one
another. All the relations of Appendix A hold with primed quantities.
We can therefore repeat word by word the derivation of the sliver
from eq.(\ref{Xi}) through eq.(\ref{ener}). The new solution will
have the form (\ref{Xi}) with $S$ along the parallel directions and $S$
replaced by $S'$ along the perpendicular ones. In turn $S'$ is
obtained as a solution to eq.(\ref{SS}) where all the quantities are
replaced by primed ones. This amounts to solving eq.(\ref{algeq}) with
primed quantities. Therefore in the transverse directions $S$ is replaced
by $S'$, given by
\beq
S'=CT',\quad\quad T'= \frac 1{2X'} (1+X'-\sqrt{(1+3X')(1-X')})\label{lump}
\eeq
In a similar way we have to adapt the normalization and energy formulas
(\ref{norm},\ref{ener}).

Exactly as in the sliver case, we can consider the solution with $T'$
replaced by $1/T'$. The same considerations hold as in that case.

\newsection{Spectroscopy and diagonal representation}

The diagonalization of the $X$ matrix was carried out in \cite{RSZ5},
while the same analysis for $X'$ was accomplished in \cite{Feng} and
\cite{belov1}. Here, for later use, we summarize the results of
these references. The eigenvalues of $X=X^{11},X_+=X^{12},X_-=X^{21}$
and $T$ are given, respectively, by
\be
&& \mu^{rs}(k) = \frac{1-2\,\delta_{r,s}+e^{\frac{\pi k}2}\, \delta_{r+1,s}
+ e^{-\frac{\pi k}2}\, \delta_{r,s+1}}
{1+2\, {\rm cosh} {\frac{\pi k}2} }\label{muspec}\\
&&t(k) = - e^{-\frac {\pi |k|}2}\label{tspec}
\ee
where $-\infty<k<\infty$. The generating function for the eigenvectors is
\be
f^{(k)}(z) = \sum_{n=1}^\infty v_n^{(k)} \frac {z^n }{\sqrt{n}} =
\frac 1k (1-e^{-k\,{\rm arctan}\, z}) \label{fk}
\ee
The completeness and orthonormality equations for the  eigenfunctions
are as follows
\be
\sum_{n=1}^\infty v_n^{(k)}v_n^{(k')} = {\cal N}(k) \delta(k-k'),
\quad\quad {\cal N}(k) = \frac 2k \,{\rm sinh} \frac {\pi k}2,\quad\quad
 \int_{-\infty}^{\infty} dk\,\frac{v_n^{(k)} v_m^{(k)}}{{\cal N}(k)} =
\delta_{nm}\label{orth}
\ee

The spectrum of $X$ is continuous and lies in the interval $[-1/3,0)$. It is
doubly degenerate except at $-\frac 13$. The continuous spectrum of $X'$ lies in the
same interval, but $X'$ in addition has a discrete spectrum. To describe it
we follow \cite{belov1}. We consider the decomposition
\beq
X^{'rs} = \frac 13 (1+\a^{s-r} CU' + \a^{r-s} U'C )\label{decomp}
\eeq
where $\a=e^{\frac {2\pi i}3}$. It is convenient to express everything in terms of $CU'$
eigenvalues and eigenvectors (see Appendix B). The discrete
eigenvalues are denoted by $\xi$
and $\bar \xi$. Since $CU'$ is unitary they lie on the unit circle. They are
more effectively represented via the parameter $\eta$, (\ref{xi}), which
in turn is connected to the parameter $b$ (\ref{eigeneq}). To each value
of $b$ there corresponds a couple of values of $\eta$ with opposite sign
(except for $b=0$ which implies $\eta=0$).

The eigenvectors corresponding to the continuous spectrum are
$V_N^{(k)}$  ($-\infty<k<\infty$), while the eigenvectors of the discrete
spectrum are denoted by  $V_N^{(\xi)}$ and  $V_N^{(\bar\xi)}$ .
They form a complete basis.
They will be normalized so that the completeness relation takes the form
\be
\int^{\infty}_{-\infty} dk\, V_N^{(k)}V_M^{(k)} + V_N^{(\xi)}V_M^{(\xi)}
+V_N^{(\bar \xi)}V_M^{(\bar \xi)} = \delta_{NM}\label{complete}
\ee

It has become familiar and very useful to expand all the relevant quantities
in VSFT by means of this basis. To this end we define
\be
&&a_k = \sum_{N=0}^\infty V_N^{(k)} a_N,\quad\quad a_\xi=\sum_{N=0}^\infty
V_N^{(\xi)}a_N ,\quad\quad a_{\bar\xi}=\sum_{N=0}^\infty
V_N^{(\bar\xi)}a_N \0\\
&&a_N= \int_{-\infty}^\infty dk\,V_N^{(k)} \,a_k +V_N^{(\xi)}a_\xi +
V_N^{(\bar\xi)} a_{\bar\xi}\label{abasis}
\ee
and introduce the even and odd twist combinations
\be
e_k= \frac{a_k+Ca_k}{\sqrt{2}},\quad\quad e_\eta= \frac{a_\xi+Ca_\xi}{\sqrt{2}},
\quad\quad
o_k= \frac{a_k-Ca_k}{i\sqrt{2}},\quad\quad o_\eta= \frac{a_\xi-Ca_\xi}{i\sqrt{2}},
\label{ekok}
\ee
The commutation relations among them are
\be
[e_k,e^\dagger_{k'}]=\delta(k-k'),\quad\quad
[e_\eta,e_\eta^\dagger]=1,\quad\quad
[o_k,o^\dagger_{k'}]=\delta(k-k'),\quad\quad [o_\eta,o_\eta^\dagger]=1,\0\\
\ee
while all the other commutators vanish. The twist properties are defined by
\be
Ca_k=a_{-k},\quad\quad Ca_\xi=a_{\bar\xi},\0
\ee

Using these combinations the three--strings vertex can be cast in diagonal
form and, for instance, the exponent of the conventional lump state can
be written
\be
a^\dagger S' a^\dagger &\!=\!& \int_{-\infty}^\infty dk\,t(k)\,
(a_k^\dagger,Ca_k^\dagger) + 2 \,t_\xi\, (a_\xi^\dagger, Ca_\xi^\dagger)\0\\
&\!=\!&
\frac 12\int_{-\infty}^\infty dk\,t(k)\,
 (e_k^\dagger e_k^\dagger+o_k^\dagger o_k^\dagger)+  \,t_\eta\,
  (e_\eta^\dagger e_\eta^\dagger+o_\eta^\dagger o_\eta^\dagger)
\label{diagT}
\ee
where $t_\eta\equiv t_\xi=e^{-|\eta|}$.
The unconventional lump is obtained by replacing $t_\eta$
with its inverse $e^{|\eta|}$.

In the sequel we need the behaviour of the eigenvectors when $b\to 0$ and
when $b\to \infty$. Near $b=0$ we have
\be
&& b \approx 0,\quad\quad \eta\approx 0,\quad\quad \xi\approx 1\0\\
&& V_0^{(\xi)}= \frac 1{\sqrt{2}}+ {\cal O}(\eta),\quad\quad
V_n^{(\xi)}= {\cal O}(\eta^2)\label{bnear0}
\ee
The same behaviour holds for the $V^{(\bar\xi)}$ basis.

When $b\to \infty$ we have instead
\be
&&b\to \infty,\quad\quad b \approx 4\, \log \eta,\quad\quad \xi\approx -
e^{\frac { \pi i}{3}}\0\\
&&V_0^{(\xi)} \approx e^{-\frac {\eta}2} \sqrt{2\eta \log \eta},
\quad\quad V_n^{(\xi)} \sim e^{-\frac {\eta}2}\sqrt{\eta}\label{btoinf}
\ee
and the same for $V^{(\bar\xi)}$.

These asymptotic behaviours will be used to evaluate matrix elements such as
(\ref{S'00}). In this regard they are completely reliable (and, in any case,
backed up by numerical evidence). If we consider instead the corresponding
asymptotic expansions for the $V^{(k)}$ basis, we have to be more careful.
The point is that the expression $(V_0^{(k)})^2$, see (\ref{V0k}), would
superficially seem to vanish
in the limit $b\to \infty$, but it is in fact a representation of the
Dirac delta function $\delta(k)$, see Appendix D. Therefore the result of
taking the $b\to\infty$
limit in an integral containing $(V_0^{(k)})^2$ is to concentrate it at the point
$k=0$. This renders the generating function (\ref{genfunc}) very singular and,
consequently, such integrals as $\int dk\, V_n^{(k)}V_m^{(k)}f(k)$
must be handled with care.
As for the limit of the continuous basis when
$b\to 0$, one can see that $V_0^{(k)}\to 0$, while the other eigenfunctions
have a nonvanishing finite limit.

\newsection{Time dependent solutions: dead ends}

In order to appreciate the very nature of the problem of finding
time--localized VSFT solutions, let us examine first some
obvious attempts and learn from their failure. The first thing
that comes to one's mind is to start from a lump with one transverse
space direction (therefore it represents a D24-brane) and inverse--Wick--rotate it.
One such solution has been introduced above, see section 2 from
eq.(\ref{split}) through eq.(\ref{lump}). For simplicity we denote
the transverse direction coordinate, momentum and oscillators simply by
$x,p$ and $a_N$. The solution is written as follows:
\be
&& \ket{\Psi'} = \ket {\Xi}_{25} \otimes \ket{\Lambda'}\0\\
&& \ket{\Lambda'}=  {\cal N}'\, \exp\left[{-\frac 12 \sum_{N,M\geq
0} a_N^\dagger S'_{NM} a_M^\dagger}\right]
\ket{\Omega_b}\label{1dlump} \ee
where $\ket{\Xi}_{25}$ is the
usual sliver along the longitudinal 25 directions and
\beq
{\cal N}' = \sqrt{3} \frac {V_{00}+ \frac b2}{(2\pi b^3)^\frac 14}
\sqrt{\det(1-X')\det(1+T')}\label{norm'}
\eeq
In order to study the space profile of this solution in the transverse direction we
contract it with the $x_0$--coordinate eigenstate
\be
\ket{x_0}= \left(\frac 2{b \pi}\right)^{\frac 14 } \exp \left[ -\frac 1b x_0^2 - \frac
2{\sqrt{b}}i a_0^\dagger x_0 +\frac 12
(a_0^\dagger)^2\right]\ket{\Omega_b}\label{xeigen}
\ee
The result is
\be
\bra {x_0} {\Lambda'}\rangle=\left(\frac 2{b \pi}\right)^{\frac 14 } \frac
{{\cal N}'}{\sqrt{1+s'}}\exp\left[ \frac 1b \frac{s'-1}{s'+1} x_0^2 -
\frac {2i}{\sqrt{b}}  \frac {x_0 f_0}{1+s'}-\frac 12 a^\dagger W'
a^\dagger\right]\label{xlambda}
\ee
 where the condensed notation means
 \be
 f_0 = \sum_{n=1}S'_{0n}a_n^\dagger,\quad\quad a^\dagger
W'a^\dagger= \sum_{n,m=1}a_n^\dagger W'_{nm}a^\dagger_m,\quad\quad
W'_{nm} = S'_{nm}- \frac{S'_{0n}S'_{0m}}{1+s'}\label{t0W'}
\ee
and
\beq
s'= S_{00}'\label{s'}
\eeq
After an inverse Wick--rotation
$x_0\to i{\rm x}_0, a_n^\dagger \to i a_n^\dagger$ (\ref{xlambda})
becomes
\be
\bra {{\rm x}_0} {\Lambda'}\rangle=\left(\frac 2{b
\pi}\right)^{\frac 14} \frac {{\cal N}'}{\sqrt{1+s'}} \exp\left[\frac 1b
\frac{1-s'}{1+s'}{\rm x}_0 ^2 + \frac {2i}{\sqrt{b}} \frac  {{\rm
x}_0 f_0}{1+s'}+\frac 12 a^\dagger W'
a^\dagger\right]\label{tlambda}
 \ee
  We are interested in solutions
localized in time. The second term in the exponent gives rise to
time oscillations. Only the first term  can guarantee time
localization. Precisely this happens when $|s'|>1$. However such a
condition can never be achieved within the present scheme in which
ordinary lump solutions are utilized. In fact it is possible to
show that for such solutions $|s'|\leq 1$. Therefore with the
simple--minded scheme considered so far it is impossible to
achieve time localization (in this regard our negative conclusion
is similar to \cite{hata}; as for the case $b\to 0$, see below).

Let us see this in more detail by showing that $|s'|\leq 1$. Using
the basis of the  previous section we can write \be s'\equiv
S'_{00} = \int_{-\infty}^{\infty} dk\, V_0^{(k)}(-e^{-\frac {\pi
|k|}2}) V_0^{(k)}+ V_0^{(\xi)}e^{-|\eta|}V_0^{(\xi)}  +
V_0^{(\bar\xi)}e^{-|\eta|}V_0^{(\bar\xi)}   \label{S'00} \ee Using
(\ref{V0k}), one can see that the first term in the RHS does not
contribute in the limit $b\to 0$ (i.e. $\eta\to 0$) and using the
approximants (\ref{bnear0}) we immediately see that the remaining
two terms add up to 1. Therefore when $b\to 0$, $s'\to 1$.
Viceversa, in the limit $b\to\infty$, using (\ref{btoinf}) we see
that the last two terms in the RHS of (\ref{S'00}) do not
contribute, while the first term contribute exactly --1. This can
be also shown numerically or with the alternative analytical
method of Appendix C. For generic values of $b$ we cannot
calculate $s'$ analytically but it is easy to evaluate it
numerically and to show that it is a monotonically decreasing
function of $b$ for $0\leq b<\infty$. This in turn implies that
the quantity $\frac{1-s'}{1+s'}$ is always {\it positive} (see
figure 1).
\begin{figure}[htbp]
    \hspace{-0.5cm}
\begin{center}
    \includegraphics[scale=1]{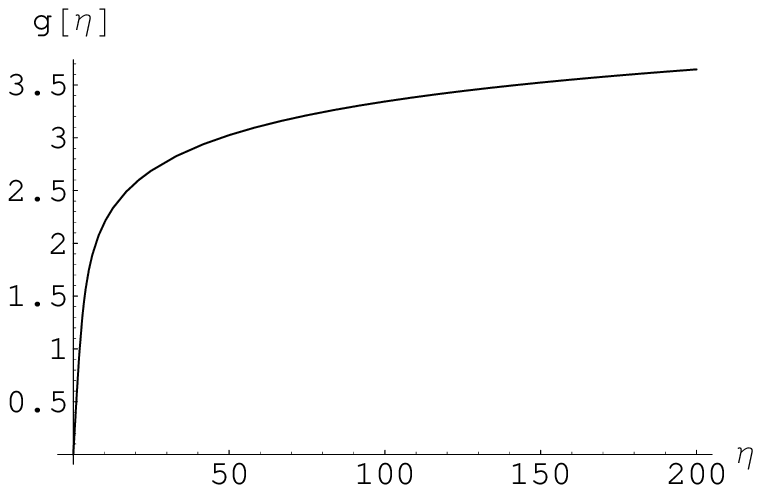}
    \end{center}
\caption{\emph{\small The quantity $g[\eta]=\frac{1-
s'}{1+s'}$ as a function of $\eta$}}
    \label{fig:A}
\end{figure}
\\

Our conclusion is therefore that we cannot obtain a time--localized solution
by inverse--Wick--rotating an ordinary lump solution.
We will show elsewhere that
also adding a constant $B$--field does not change this negative conclusion.
Some drastic change has to be made in order to produce a localized
time--dependent solution.

\newsection{A Rolling Tachyon--like Solution}

It is not hard to realize that if we were to replace $e^{-|\eta|}$ with
$e^{|\eta|}$ in eq.(\ref{S'00}) we would reverse the conclusion at the
end of the previous section. In fact, see below, we would have $|s'|\geq 1$.
In this section we wish to exploit this possibility. In section 2 we have
seen that if in the lump solution we replace $T'$ by $1/T'$, formally,
we still have a projector.
Motivated by this fact we define an unconventional lump, by replacing
$\ket{\check\Lambda'}$ in (\ref{1dlump}) with
\beq\label{rollsliv}
\ket{\check\Lambda'}=\check\N' \exp\left(-\frac12 a^\dagger C {\check T}' a^\dagger\right)
\ket{\Omega_b}
\eeq
where
\beq
{\check T}'_{NM}=-\int_{-\infty}^\infty dk \,V^{(k)}_N\, V^{(k)}_M
\exp\left(-\frac{\pi|k|}{2}\right)+
\left(V_N^{(\xi)}\,V_M^{(\xi)}+V_N^{(\bar\xi)}\,
V_M^{(\bar\xi)}\right)\exp\,|\eta|\label{T'NM}
\eeq
Due to the fact that the star product is split into eigenspaces
of the Neumann coefficients $X',\, X_+',\, X_-'$, the projector equation split
accordingly into the continuous and discrete spectrum part.
Therefore we are guaranteed that (\ref{rollsliv}) is again a projector,
as one can
on the other hand easily verify by direct calculation.
This is the solution we propose.

Before we proceed with our analysis we would like to clarify a basic question
about the solution we have just put forward. Passing from
a squeezed state solution with a matrix $T'$ to another characterized
by the inverse matrix $1/T'$ may lead in general to unacceptable
features of the state, such as divergent terms in the oscillator basis.
However in the case at hand, in which one inverts
only the discrete spectrum, such unpleasant aspects disappear. First of all
the matrix ${\check T}'$ is well defined both in the oscillator and in
the diagonal
basis. Second, such expression as $\sqrt{\det(1-{\check T}')}$ are well--defined.
This is due to the fact that, if we are allowed to factorize the discrete
and continuous spectrum contribution, the former can be written as
$ \det(1-{\check T}')^{(d)} = (1-\exp |\eta|)^2$, so that the possible
dangerous -- sign under the square root disappears due to the double multiplicity
of the discrete eigenvalue. Third, the energy density of the
(Euclidean ) solution (\ref{rollsliv}) equals the energy
density of the ordinary lump. In fact, using the formulas of \cite{RSZ2},
the ratio between the energy densities of the two solutions reduces to
\be
\sqrt{\frac {\det(1+{\check T}')}{\det(1-{\check T}')}}/
\sqrt{\frac {\det(1+ T')}{\det(1- T')}} = \sqrt{\frac {(1+e^{|\eta|})^2}
{(1-e^{|\eta|})^2}}/\sqrt{\frac{(1+e^{-|\eta|})^2}{(1-e^{-|\eta|})^2}}=1\label{ratio}
\ee
after factorization of the discrete and continuous parts of the spectrum.

After these important remarks it remains for us to show that this
solution has the appropriate
features to represent a rolling tachyon solution. To see if this is true
we have to represent it
in a more explicit way. In particular we have to extract the explicit time
dependence (better, the space dependence and then inverse--Wick--rotate it).
To do so, we have to choose a (coordinate) basis on which to project (\ref{rollsliv}).
There seem to be two distinguished ways to make this choice. We will work them out explicitly
and then discuss them.

To start with let us define the following coordinate and momentum operator,
given by the twist even and twist odd parts of the discrete spectrum,
\be
\hat x_\eta&=&\frac{i}{\sqrt{2}}(e_\eta-e_\eta^\dagger)\label{discrtime}\\
\hat y_\eta&=&\frac{i}{\sqrt{2}}(o_\eta-o_\eta^\dagger)\label{discrmom}
\ee
 The eigenstates of the coordinate $\hat x_\eta$ are given by
\be
&\ket{x}=\frac{1}{\sqrt{\pi}}\exp\left(-\frac 12  x^2 -\sqrt{2}i e_{\eta}^\dagger x + \frac 12
e_{\eta}^\dagger e_{\eta}^\dagger\right)\ket{\Omega_{\eta_e}},&\label{dtimestate}\\
&e_{\eta}\ket{\Omega_{\eta_e}}=0&\0\\
&\hat x_\eta\ket{x}=x\ket{x}&\0
\ee
Correspondingly the eigenstates of the momentum $\hat y_\eta$ are
\be
&\ket{y}=\frac{1}{\sqrt{\pi}}\exp\left(-\frac 12  y^2
-\sqrt{2}i o_{\eta}^\dagger  y + \frac 12
o_{\eta}^\dagger o_{\eta}^\dagger\right)\ket{\Omega_{\eta_o}},&\label{dmomstate}\\
&o_{\eta}\ket{\Omega_{\eta_o}}=0&\0\\
&\hat y_\eta\ket{y}=y\ket{y}&\0
\ee

In order to make
the $x,y$ dependence explicit we  project our solution
(\ref{rollsliv}) into the position/momentum eigenstates (\ref{dtimestate},
 \ref{dmomstate}).
Using standard results\footnote{Here we are assuming that the vacuum
factorizes into $\ket{\Omega_{\eta_e}}\otimes \ket{\Omega_{\eta_o}}\otimes
\ket{\Omega_{c}}$ where the latter factor represents the vacuum
with respect to the continuous oscillator component.} we get
\beq\label{deprofile}
\bra{x,y}\check\Lambda'\rangle=\frac{1}{\pi(1+e^{|\eta|})}
\exp\left(\frac{e^{|\eta|}-1}{e^{|\eta|}+1}
(x^2+y^2)\right)|\check\Lambda'_c\rangle
\eeq
The state $|\check\Lambda'_c\rangle$ is given by (\ref{rollsliv}), but
with only oscillators from the continuous spectrum, as the contribution of
the discrete spectrum is now contained in the prefactor at the rhs of (\ref{deprofile}) .
Now we perform the inverse Wick rotation $x\to i{\rm x}$, $y\to -i{\rm y}$
to recover the Lorentz signature, and obtain
\beq\label{dmprofile}
|\check\Lambda'
({\rm x} ;{\rm y})\rangle=\frac{1}{\pi(1+e^{|\eta|})}
\exp\left(-\frac{e^{|\eta|}-1}{e^{|\eta|}+1}
({\rm x}^2+{\rm y}^2)\right)|\check\Lambda'_c\rangle^{(Wick)}
\eeq
It is evident that for every value of $\eta$ the solution is localized in
the ${\rm x}$--time coordinate. The extra coordinate ${\rm y}$ is related to
internal twist odd degrees of freedom and can be interpreted as  a free parameter
of the representation (\ref{dmprofile}).
This solution also contains the free parameter $\eta$ which is
nothing but a reparametrization of $b$, through (\ref{eigeneq}). Therefore
it is characterized by two free parameters.

The `time' ${\rm x}$ is not the ordinary time, i.e. the time coupled to the
flat open string metric and related to the string center of mass.
We will see later on a possible interpretation for ${\rm x}$. Now,
let us turn to the ordinary (open string) time, i.e. the time defined
by the center of mass of the string and analyze the corresponding
time profile. Despite the fact that this coordinate  is not diagonal
for the $*$--product
we can  still have complete control on the profile
along it. The center of mass position operator is given by
\beq\label{centmass}
\hat x_0=\frac{i}{\sqrt{b}}(a_0-a_0^\dagger)
\eeq
The center of mass position eigenstate is
\beq\label{centstate}
\ket{x_0}=\left(\frac{2}{b\pi}\right)^{\frac14}\exp\left( -\frac 1b x_0x_0 -\frac{2}{\sqrt{b}}i
a_0^\dagger
x_0+\frac{1}{2} a_0^\dagger a_0^\dagger\right)\ket{\Omega_b}
\eeq
Let us compute the center of mass time profile. After inverse--Wick--rotating it,
it turns out to be
\be
\ket{\check\Lambda'({\rm x}_0)} &\!=\!& \bra{{\rm x}_0}\check\Lambda'\rangle =
\label{centimepro1}\\
\left(\frac{2}{b\pi}\right)^{\frac14}\frac{\check\N'}{\sqrt{1+{\check T}'_{00}}}&&\!
\exp\left(\frac1b\frac{1-{\check T}'_{00}}{1+{\check T}'_{00}}{\rm x}_0^2+
\frac{2i}{\sqrt{b}(1+{\check T}'_{00})}{\rm x}_0 {\check T}'_{0n}a_n^\dagger
+\frac 12 a_n^\dagger W'_{nm} a_m^\dagger\right)\ket{\Omega_b}\0\\
 W'_{nm}&\!=\!& {\check S}'_{nm}-
 \frac{1}{1+{\check T}'_{00}}{\check S}'_{0n}{\check S}'_{0m}\label{centimepro2}
\ee
The quantities ${\check S}'_{0n}$ and ${\check S}'_{nm}$ can be computed in the diagonal basis
\be
{\check S}'_{0n}&=&{\check T}'_{0n}\label{s'0n}\\
       &=&(1+(-1)^n)\left(-\int_0^\infty V_0^{(k)} V_n^{(k)}
\exp\left(-\frac{\pi k}{2}\right)+V_0^{(\xi)}\,V_n^{(\xi)}\,\exp\,|\eta|\right)\0\\
{\check S}'_{nm}&=&(-1)^n {\check T}'_{nm}=\label{s'nm}\\
       &=&((-1)^n+(-1)^m)\left(-\int_0^\infty V_n^{(k)} V_m^{(k)}
\exp\left(-\frac{\pi k}{2}\right)+V_n^{(\xi)}V_m^{(\xi)}\,\exp\,|\eta|\right)\0
\ee

It is evident that the leading time dependence in (\ref{centimepro1}),
for large ${\rm x}_0$, is contained in
$\exp\left(\frac1b\frac{1-{\check T}'_{00}}
{1+{\check T}'_{00}}{\rm x}_0^2\right)$. The number ${\check T}'_{00}$ is
$b(\eta)$--dependent and can be computed  via
\beq\label{T'00}
{\check T}'_{00}(\eta)=-2\int_0^\infty dk \left(V_0^{(k)}(b(\eta))\right)^2
\exp\left(-\frac{\pi k}{2}\right)+2({V_0^{(\xi)}})^2\exp\,|\eta|
\eeq
This is the crucial quantity as far as the time profile is concerned.
An analytic evaluation of it is beyond our reach. However we will later show
that
\be
\lim_{\eta\to 0}{\check T}'_{00}&=&1\label{T'00eta0}\\
\lim_{\eta\to \infty}{\check T}'_{00}&=&\infty\label{T'00etainfty}
\ee
A numerical analysis shows that this quantity is a function
monotonically increasing with $\eta$ within such limits. This
means that the quantity $\frac{1-{\check T}'_{00}}{1+{\check
T}'_{00}}$ is always {\it negative} (it lies in the interval
$(-1,0)$, see figure 2) and so the profile is always localized in the center of
mass time, except in the extreme case $\eta\to 0$, which
corresponds to the tensionless limit.
\begin{figure}[htbp]
    \hspace{-0.5cm}
\begin{center}
    \includegraphics[scale=1]{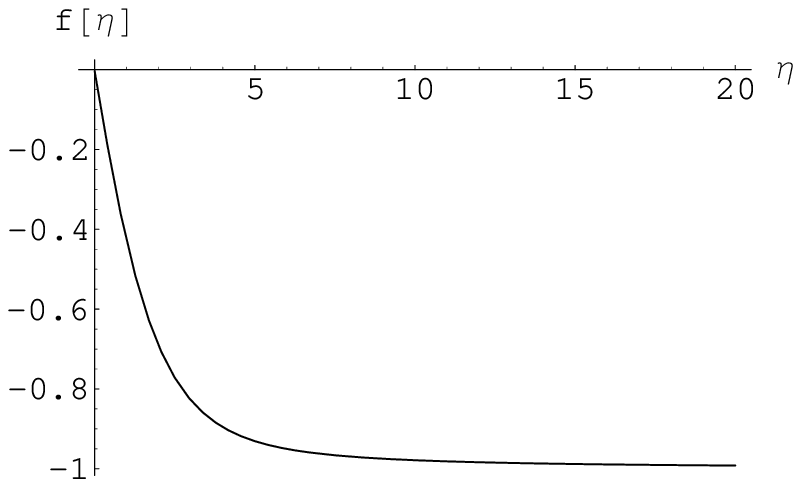}
    \end{center}
\caption{\emph{\small The quantity $f[\eta]=\frac{1-{\check
T}'_{00}}{1+{\check T}'_{00}}$ as a function of $\eta$}}
    \label{fig:B}
\end{figure}

This has to be compared with the usual lump solution (see previous section) for
which the corresponding quantity is always positive and takes
values in the interval $(0,\infty)$, allowing for localized {\it
space} profiles but divergent along a timelike direction.

For reasons that will become clear in the next section, we extract also
the free parameter ${\rm y}$ dependence, by
projecting onto the corresponding twist--odd eigenstate (\ref{dmomstate}).
This operation
can be done before or after the projection along the center
of mass coordinate and does not interfere with it because $\hat{\rm y}$ does not
contain the zero mode.
We will therefore consider the following representation
of our solution (inverse Wick rotation is again understood)
\be
\ket{\Lambda'({\rm x}_0,y)}&\!=\!&\bra{{\rm x}_0,{\rm y}}\check\Lambda'\rangle=
\left(\frac{2}{b\pi}\right)^{\frac14}\frac{\check\N'}{\sqrt{2\pi(1+e^{|\eta|})}}
\exp\left(\frac{1-e^{|\eta|}}{1+e^{|\eta|}}{\rm y}^2\right)\label{x0yprof}\\
\cdot\frac{1}{\sqrt{1+{\check T}'_{00}}}\!\!&&\!\!
\exp\left(\frac1b\frac{1-{\check T}'_{00}}{1+{\check T}'_{00}}{\rm x}_0^2+
\frac{2i}{\sqrt{b}(1+{\check T}'_{00})}{\rm x}_0 {\check T}'_{0n}a_n^\dagger-\frac12
a_n^\dagger W''_{nm} a_m^\dagger\right)\ket{0}\0
\ee
The quantities ${\check T}'_{00}$ and ${\check T}'_{0n}$ are the same as in
(\ref{T'00}, \ref{s'0n}) since the momentum $\hat y_\eta$ is twist--odd.
Some changes occur in
$W''_{nm}$
\be
W''_{nm}&=&{\check S}''_{nm}-\frac{1}{1+{\check T}'_{00}}
{\check S}'_{0n}{\check S}'_{0m}\label{W''nm}\\
{\check S}''_{nm}&=&((-1)^n+(-1)^m)\left(-\int_0^\infty dk\,V_n^{(k)}\,
V_m^{(k)}\,\exp\left(-\frac{\pi k}{2}\right)+V_n^{(\eta)}V_m^{(\eta)}
\exp\,|\eta|\right)\quad n,m\;{\rm even}\0\\
        &=&-((-1)^n+(-1)^m)\int_0^\infty dk\, V_n^{(k)}\,
    V_m^{(k)}\,\exp\left(-\frac{\pi k}{2}\right)\quad n,m\;{\rm odd}\0
\ee
Note that ${\check S}''_{nm}$ gets contribution only from the twist--even part
of the discrete spectrum.

In conclusion (\ref{x0yprof}) provides the solution we were looking for.
It represents a solution localized in ${\rm x_0}$, with the desired
profile. It depends on two free parameters ${\rm y}$ and
$\eta$ (or $b$). These are all positive features. But let us start
making a closer comparison with the rolling tachyon solution (such a comparison
is made with the representation (\ref{x0yprof})).
This can be done by considering the limit $b\to \infty$,
which can be derived from the
eqs.(\ref{btoinf}). $b\to\infty$ means $\eta\to\infty$ (for simplicity
from now on we take $\eta$ positive) and
\be
\check T'_{00} \approx 2\, \eta\,\log \,\eta\, (1 - \frac {\log (2\pi)}{\log\,{\eta}}+
\ldots)\label{s'approx}
\ee
where dots denote higher order terms.
Therefore we see that in this limit any time dependence in (\ref{x0yprof})
disappears. Moreover, anticipating a result of the following section,
we also have that $W''_{nm} \to S_{nm}$.
In other words, in the limit $b\to\infty$ we obtain a static solution
corresponding to the initial sliver. From this we understand that the
parameter $1/b$, for large $b$, plays a role similar to Sen's parameter
$\tilde\lambda$ near 0\footnote{
We recall that Sen's rolling tachyon solution  depends
on the parameter  $\tilde\lambda$,
which appears in the
$\tilde \lambda\int_{\partial D} dt\cosh X^0(t)$--deformed BCFT.}.
A second remark concerns the limit $y\to \infty$.
In this case the first exponential factor in the RHS of (\ref{x0yprof})
suppresses everything, so that the limit is the 0 state. In other words,
we can consider this value of the parameter $y$ as identifying the (relatively)
stable vacuum state.

Although the rolling tachyon naturally compares with (\ref{x0yprof})
rather than with (\ref{dmprofile}), it is instructive to
repeat something similar with the latter. Let us stress once more that both
(\ref{x0yprof}) and (\ref{dmprofile}) represent the same solution, but
in different bases, in particular with two different times: one, ${\rm x}_0$,
is the open string center of mass time, the other, ${\rm x}$. is related to the
discrete spectrum.
In the (\ref{dmprofile}) case a parameter like $b$ is missing.
But this is something that is simply not customary and can be
easily remedied. We can in fact introduce
a parameter $b_e$ in (\ref{discrtime},\ref{discrmom}), just replacing $\sqrt{2}$
with $\sqrt{b_e}$ in those equations. Then (\ref{dmprofile}) would become
\beq\label{dmprofileb}
|\check\Lambda'
({\rm x} ;{\rm y})\rangle=\frac{1}{b_e\pi(1+e^\eta)}
\exp\left(-\frac 1{b_e}\frac{e^\eta-1}{e^\eta+1}
({\rm x}^2+{\rm y}^2)\right)|\check\Lambda'_c\rangle^{(Wick)}
\eeq
and we could repeat the same argument as above and reach the same
conclusion, except that in this case we have to take $b_e\to \infty$
as well as $\eta\to\infty$. The limit $y\to\infty$ plays the same
role as in the (\ref{x0yprof}) representation.

In the next section we will study the solution (\ref{rollsliv}) in a
regime we are more familiar with, the low energy regime $\a'\to 0$, and
in the other extreme regime, $\a'\to\infty$, in which the solution considerably
simplifies and an analytic treatment is possible. What we would like to
see more closely is whether, for
sufficiently small values of the parameters, the solution at time 0
is close enough to the sliver configuration (\ref{Xi}), whose decay the
solution is expected to describe.

\newsection{Low energy and tensionless limits}

As reviewed in appendix C, the low energy limit is
obtained by performing an $\epsilon\to 0$ limit on the quantities
that depend on the Neumann coefficients of the three strings vertex.
$\epsilon$ is a dimensionless parameter that represents the smallness
of $\a'$, \cite{MT}. As it happens, in all the expansions we consider,
the parameters $\e$ and $b$ only appear through the ratio $\e/b$.
Therefore, formally, the expansions for small $\e/b$ are the same as the
expansions for large $b$, i.e. $\eta\to \infty$. Therefore, in this section,
when we consider the expansion in $\eta$ near $\infty$
we really mean the expansion for $\e/b$ small (i.e. $\e$ small and $b$ finite).
A different attitude is required by the `external' states like
(\ref{xeigen}). There the rescaling of $x_0$ would lead to the
replacement $b \to b\e$. In this case we absorb $\e$ into $x_0$ and keep $b$
finite.
In conclusion, throughout the analysis of the low energy limit,
$b$ should be considered as a finite free parameter.

Let us analyze in detail what is the limit of the various quantities appearing
in (\ref{x0yprof}). First of all we have
\beq\label{lowcoeff}
\lim_{\eta\to\infty}\frac{1-\check T'_{00}}{1+{\check T}'_{00}}=-1
\eeq
This follows from (\ref{btoinf}) and from the discussion at the end of
section 3, in particular from the property of $(V_0^{(k)})^2$ of approximating
$\delta(k)$ in the limit $b\to \infty$, which implies that ${\check T}'_{00}\to\infty$
in the same limit. For the oscillating term we have
\beq\label{lowosc}
\lim_{\eta\to\infty}\frac{{\check T}'_{0n}}{1+{\check T}'_{00}}=\lim_{\eta\to\infty}
\frac{1}{\sqrt{2\,\log\,\eta}}=0
\eeq
To evaluate this limit one must evaluate ${\check T}'_{0n}$. This in turn
requires knowing the asymptotic expansion of the basis $V_n^{(k)}$
for $\eta\to\infty$. This is done in Appendix D. A numerical approximation
confirms the above result.

Thus, in the limit, the oscillating part completely decouples from the time dependent part.
It remains for us to consider the limit of the quadratic form $W''_{nm}$, (\ref{W''nm}).
When $n,m$ are odd there are no contributions from the discrete
spectrum, since the contraction with $\bra{\rm y}$ has eliminated them.
\beq\label{lowodd}
W''_{2n-1,2m-1}={\check S}^{'(c)}_{2n-1,2m-1},
\eeq

When $n,m$ are even we have, on the contrary, potentially dangerous terms
because there are divergent contributions arising from the
discrete spectrum. The latter have to be carefully evaluated.
\be
W^{''(d)}_{2n,2m}&=& 2V_{2n}^{(\xi)}V_{2m}^{(\xi)}\left(e^\eta-
\frac{2 ({V_0^{(\xi)}})^2 e^{2\eta}}{2 ({V_0^{(\xi)}})^2 e^{\eta}+
{\cal O}(\frac{e^{-\eta}}{\eta \,\log \eta})}\right)\0\\
&&=2V_{2n}^{(\xi)}V_{2m}^{(\xi)}\,{\cal O}(\frac{e^{-\eta}}{\eta
\,\log \eta}) \approx {\cal O}(\frac 1{\log \eta}) \label{loweven}
\ee We see that the potentially divergent contributions arising
from the discrete spectrum exactly cancel when $\eta\to\infty$.
Therefore, as far as $W^{''}_{2n,2m}$ is concerned, we are left
only with the contribution from the continuous spectrum. Of the
two pieces that contribute to $W^{''(c)}_{2n,2m}$, see
eq.(\ref{W''nm}) only the first survives in the limit
$\eta\to\infty$, the second vanishes for the usual reasons.
Therefore we can conclude that \be W''_{nm} = {\check
S}^{'(c)}_{nm}+...\0 \ee where dots denote subleading corrections
of order at least $1/\log\,\eta$. At this stage we can do the
calculation directly as in Appendix D, or we can resort to an
indirect argument by noticing that ${\check S}^{'(c)}_{nm}$
approaches $S_{nm}^{'}$ in the same limit, because the discrete
spectrum contribution to the latter vanishes, and then use the
results of Appendix C. In both cases we  conclude that
\be
W''_{nm} = S_{nm} + {\cal O}(\e/b)\label{sliverapprox}
\ee
Going back to equation (\ref{x0yprof}) we see that, modulo a
normalization factor, we obtain
\be
\lim_{\alpha'\to 0}\ket{\check\Lambda'({\rm x}_0,{\rm y})}=
\check\N'({\rm y})\, e^{-\frac{\rm
x_0^2}{b}}\,\ket{\Xi}\label{a'to0} \ee where $\ket{\Xi}$ is the
zero momentum sliver state. This result can be phrased as follows:
in the low energy limit the solution takes the form of a
time--Gaussian multiplying a sliver, the subleading terms being
proportional to  $\e/b$, eq.(\ref{sliverapprox}).

To end this section let us briefly consider the opposite limit,
that is $\alpha'\to\infty$ (tensionless limit).
As in the previous case this is formally achieved by
taking the $\eta\to 0$ limit in all the quantities which are related to the Neumann
coefficients, but leaving $b$ as a free parameter. This limit is well
defined.
Using the results of appendix C we get
\beq\label{highcoeff}
\lim_{\eta\to 0}\frac{1-{\check T}'_{00}}{1+{\check T}'_{00}}=0
\eeq
The oscillating term in (\ref{x0yprof}) vanishes as well.
This result implies that the Gaussian representing time dependence in (\ref{x0yprof})
is actually completely flat: time dependence has disappeared!
We believe this to be related to the fact that all strings modes become massless in this
limit \cite{giulio}, so there are no modes to decay into.
It is easy to see that the only non vanishing term in the exponent of (\ref{x0yprof})
is the quadratic part which gets contribution only from the
continuous spectrum (on the contrary of the $\eta\to\infty$ limit the discrete
eigenvector has only the $0$--component, while the higher components
disappear like positive powers of $1/\eta$). We remark that in the tensionless
limit the center of mass time and the ${\rm x}$ time are identified.

\newsection{Discussion}

In the last two sections we have shown that by inverting the discrete part
of the spectrum
we obtain a definite (unconventional) lump solution which, after inverse
Wick--rotation, gives rise to a time--localized state with many properties
characteristic of
the rolling tachyon solution. In the course of our exposition we have left
aside some
loose ends which we would like now to tie up or at least comment upon.

The first comment concerns normalization of the states we have come across. We have
written down throughout normalization factors in quite a formal way. We have already
recalled the fact that the sliver state and the lump state have a vanishing
normalization, but we believe these problems have to be kept separated from the
normalization of our time dependent solution. As a matter of fact
a normalization problem
appears only for the representation (\ref{x0yprof}) and in the low energy limit,
for the coefficient $\check{\cal N}'$ in
(\ref{a'to0}) diverges exponentially for $\eta\to\infty$ once all the
contributions are taken into account (this problem does not arise for the
other representation (\ref{dmprofile})). We remark however that, as was
noticed in the discussion after eq.(\ref{rollsliv}),
the energy density of the corresponding Euclidean solution is well--defined
(once the conventional lump energy density is). Therefore the exploding normalization
can only be an artifact of the representation. It means that we have to use
the parameters of the state to regulate the normalization, although it is not
clear a priori what is the right way to do it. A possibility is to use the
factor $\exp\left(\frac{1-e^{|\eta|}}{1+e^{|\eta|}}{\rm y}^2\right)$ in
(\ref{x0yprof}). Since this
vanishes for ${\rm y}$ large, we can view ${\rm y}$ as a suitable function
of $\eta$ as
$\eta\to\infty$. This can settle the problem. Other possibilities are
connected to
dressing, \cite{BMP1, BMP2}.

We would like to add a comment concerning the meaning of our
solution (\ref{rollsliv}) before inverse Wick--rotation. As we have noticed,
its profile is an inverted Gaussian that explodes at infinity.
This suggests that we can interpret it
as a D--brane located at infinity in the transverse direction, that is
at infinite imaginary time. One could
speculate this to  be linked to the D--branes at imaginary times
referred to in \cite{Senroll,GIR}.

Another important question is the number of parameters. Our solution depends on two
parameters ${\rm y}$ and $b$. One may wonder why we extracted the ${\rm y}$ dependence
from $\ref{rollsliv}$. This is indeed not a choice but a constraint. Had we not done it,
we would have found a different formula (\ref{W''nm}) in which also the $n,m$ odd part
of $S''_{nm}$ would have taken a contribution from the discrete spectrum (exactly
as the $n,m$ even part). However in the odd--odd part no such cancelation (\ref{loweven})
as in the even--even part occurs and we would find badly divergent
coefficients in $W'$. We gather that ${\rm y}$ is a genuine free parameter of the
time--dependent state. What about $b$? It was argued in \cite{RSZ2} that this
parameter represents a gauge degree of freedom. This need not be in contradiction
with the  meaning we have attributed to it in the previous sections.
We recall that in ordinary gauge theory a singular gauge transformation may
convey some physical information. Now, looking at (\ref{osc}), the values
$b=0$ and $b=\infty$ may well correspond to singular gauge transformations, and therefore
contain physical information. More generally the gauge nature of $b$ may mean that
using a different formulation one may be able to write the solution in terms of a single
physical parameter which contains the information carried by both $b$ and ${\rm y}$.

The third question we would like to address is the
relation between the two representations (\ref{dmprofile}) and
(\ref{x0yprof}). The latter is expressed in terms of the open
string center of mass ${\rm x}_0$ and its interpretation is
obvious. The interpretation of the former is less clear since the
'time' ${\rm x}$ does not have a clear connection with the open
string center of mass time. A rather bold speculation is that
${\rm x}$ be connected with the closed string time. The closed
string time couples to the closed string metric, which, in
correspondence with the D--brane, must develop a singularity (it
must be a solution of the effective low energy field theory
associated to the closed string). So the relation between the open
and the closed string time should be something like $ g_c (dt_c)^2
\sim g_o (dt_o)^2$ in the field theory limit, were $g_o=1$ and
$g_c$ becomes larger and larger near the origin. Something similar
indeed occurs between ${\rm x_0}$ and ${\rm x}$ when
$\eta\to\infty$. In fact the ratio between ${\rm x_0}$ and the
zero mode part of ${\rm x}$ decreases exponentially with $\eta$.
We notice moreover that the normalization of the representation
(\ref{dmprofile}) does not need any regularization. In other words
${\rm x}$ seems to be a smoother choice of time, with respect ${\rm x_0}$.

Next we would like to recall that recently, \cite{GrossE},
the role of the time coordinate represented by the midpoint
$X^0\left(\frac\pi 2\right)$ for causality in SFT has been emphasized.
In our VSFT solution the profile along this time turns out to be highly
singular: it is a constant infinite function (finite only at
$X^0\left(\frac\pi 2\right)=0$), the inverse Wick--rotation of the midpoint
space profile of \cite{MT}.

To conclude, in this paper we have shown that VSFT contains solutions that
describe brane decay with several features of the rolling tachyon. Of course
a lot has still to be done. Further work is necessary to show a closer
connection between our solution and the rolling tachyon. Moreover we may ask
whether one can find solutions corresponding to half--S--branes.
More important, the problems of energy conservation as well as the nature
of the matter the branes decay into should be clarified.

\begin{center}
{\bf Acknowledgments}
\end{center}
We would like to thank I.Aref'eva and L.Rastelli for useful
discussions as well as P.Prester for his collaboration in the
early stage of this research. C.M. thanks G. Bonelli for
correspondence. This research was supported by the Italian MIUR
under the program ``Teoria dei Campi, Superstringhe e Gravit\`a''
and by CAPES-Brasil as far as R.J.S.S. is concerned.

\section*{Appendix}
\appendix
\section{A collection of well--known formulae}
In this Appendix we collect some useful results and formulas involving the
matrices of the three strings vertex coefficients.

To start with, we recall that
\begin{itemize}
\item (i) $V_{nm}^{rs}$ are symmetric under simultaneous exchange of
the two couples of indices;
\item (ii) they are endowed with the property of cyclicity in the
$r,s$ indices, i.e. $V^{rs}= V^{r+1,s+1}$, where $r,s=4$ is
identified with $r,s=1$.
\end{itemize}

Next, using the twist matrix $C$  ($C_{mn}= (-1)^m \delta_{mn}$), we define
\beq
X^{rs} \equiv C V^{rs}, \quad\quad r,s=1,2,\label{EX}
\eeq
These matrices are often rewritten in the following way
$X^{11}=X,\, X^{12}=X_+,\,
X^{21}=X_-$. They commute with one another
\beq
[X^{rs}, X^{r's'}] =0, \label{commute}
\eeq
moreover
\beq
CV^{rs}= V^{sr}C ,\quad\quad CX^{rs}= X^{sr}C
\eeq
Next we quote some useful identities:
\be
&&X+ X_++ X_- = 1\0\\
&& X_+X_- = (X)^2-X\0\\
&& (X_+)^2+ (X_-)^2= 1- (X)^2\0\\
&& (X_+)^3+ (X_-)^3 = 2 (X)^3 -  3(X)^2 +1
\label{Xpower}
\ee
The same relations hold if we replace $X,X_+,X_-,T$ by $X',X'_+,X'_-,T'$,
respectively.

\section{Diagonal representation of $CU'$}

With reference to formula (\ref{decomp}), we illustrate the spectroscopy
and diagonal representation of $CU'$.
The matrix $CU'$ is hermitian, unitary and commutes with $U'C$.
The discrete eigenvalues $\xi$ and $\bar \xi$ are
determined as follows, \cite{belov1}. Let
\be
\xi = - \frac{2-{\rm cosh} \,\eta- i \sqrt{3}\, {\rm sinh}\,\eta}
{1-2 {\rm cosh} \,\eta}
\label{xi}
\ee
and
\be
F(\eta) = \psi\left(\frac 12 +\frac \eta{2 \pi i}\right) - \psi\left(\frac
12\right),
\quad\quad \psi(z)=\frac {d \log\Gamma(z)}{dz} \label{psi}
\ee
Then the eigenvalues $\xi$ and $\bar \xi$ are the solutions of
\be
\Re F(\eta)= \frac b4\label{eigeneq}
\ee
The eigenvectors $V_n^{(\xi)}$ are defined via the generating
function
\be
F^{(\xi)}(z) =\! && \sum_{n=1}^\infty V_n^{(\xi)} \frac {z^n}{\sqrt{n}}=
- \sqrt{\frac 2b} V_0^{(\xi)} \left[ \frac b4 +
\frac{\pi}{2\sqrt{3}}\, \frac{\xi-1}{\xi+1} + \log\, iz\,\right. \0\\
&&+ \left. e^{-2i(1+\frac{\eta}{\pi i}) {\rm arctan}\, z}
\Phi(e^{-4i\,{\rm arctan}\,z},1, \frac 12 +
\frac {\eta}{2 \pi i})\right]\label{genfund}
\ee
where $\Phi(x,1,y)= 1/y\,{}_2\!F_1(1,y;y+1;x)$, while
\beq
V_0^{(\xi)} = \left({\rm sinh}\,\eta \frac {\d}{\d \eta}
\left[\log \Re F(\eta)\right]\right)^{-\frac 12}\label{V0xi}
\eeq

As for the continuous spectrum, it is spanned by the variable $k$,
$-\infty<k<\infty$.
The eigenvalues of $CU'$ are given by
\be
\nu(k)= -\frac {2 +{\rm cosh}\, \frac{\pi k}{2} + i \sqrt{3}\, {\rm sinh}
\,\frac{\pi k}{2} }{1+ 2\,{\rm cosh }\, \frac{\pi k}{2} }\0
\ee
The generating function for the eigenvectors is
\be
&&F_c^{(k)}(z) = \sum_{n=1}^{\infty} V_n^{(k)} \frac {z^n}{\sqrt{n}} = V_0^{(k)}
\sqrt{\frac 2b} \left[ -\frac b4 -\left(\Re F_c(k)-\frac b4\right)
e^{-k\,{\rm arctan}\,z}-\log\, iz\right.\label{genfunc}\\
&&-\left.\left(\frac{\pi}{2\sqrt{3}}\,\frac{\nu(k)-1}{\nu(k)+1}+
\frac {2i}k\right)
+2i\,f^{(k)}(z) -\Phi(e^{-4i\,{\rm arctan}\,z},1,1  +
\frac {k}{4 i})\,e^{-4i\,{\rm arctan}\,z}\,e^{-k\,{\rm arctan}\,z}
\right]\0
\ee
where
\be
F_c(k) = \psi(1 +\frac k{4 \pi i}) - \psi(\frac 12)\0,
\ee
while
\be
V_0^{(k)} = \sqrt{\frac b{2{\cal N}(k)}}
\left[4 +k^2\left(\Re F_c(k)-\frac b4\right)^2\right]^{-\frac 12}
\label{V0k}
\ee

The continuous eigenvalues of $X',X_-',X_-'$ and $T'$ (for the conventional
lump) are given by same
formulas as for the $X,X_+,X_-$ and $T$ case, eqs(\ref{muspec},\ref{tspec}).
As for the discrete eigenvalues, they are given by the formulas
\be
&& \mu^{rs}_\xi =\frac {1-2\,\delta_{r,s}-e^\eta\,
\delta_{r+1,s}-e^{-\eta}\,\delta_{r,s+1}}{1-2\,{\rm cosh}\,\eta} \0\\
&&t_\xi= e^{-|\eta|}\label{xtspec}
\ee

\section{Limits of $X'$ and $T'$}

In this Appendix we briefly discuss the low energy and high energy limit of $X'$
and $T'$ in the oscillator basis.
The Neumann coefficients $V_{NM}^{'(rs)}$ we use are given in Appendix B
of \cite{RSZ2}. They explicitly depend on the $b$ parameter.
In the low energy limit the three--strings vertex can be expanded by
means of a parameter $\e$ (a dimensionless parameter, in fact an alias of
$\a'$) , see \cite{MT}. This translates into an expansion for
$V_{NM}^{'(rs)}$ triggered by the following rescalings
\be
&& V^{(rs)}_{mn}  \rightarrow  V_{mn}^{(rs)} \0  \\
&& V_{m0}^{(rs)}  \rightarrow  \sqrt{\epsilon} V_{m0}^{(rs)} \label{Ve} \\
&& V_{00}  \rightarrow  \epsilon V_{00} \0
\ee
For instance $X'$ is expanded as follows to the lowest orders of approximation
\be
X' = \left(\matrix{-\frac 13 +\frac 83 {V_{00}}\frac {\e}{b}& -\frac 43
\sqrt{\frac {2\e}b } \bra {{\bf{v}}_e}\cr
-\frac 43
\sqrt{\frac {2\e}b } \ket {{\bf{v}}_e}&
X- \frac 83 \frac {\e}b ( \ket {{\bf{v}}_e} \bra {{\bf{v}}_e}
-\ket {{\bf{v}}_o} \bra {{\bf{v}}_o}}\right)\label{X'approx}
\ee
where
\be
\ket {{\bf{v}}_e}_n = -\frac 3{2\sqrt{2}} V_{0n}^{(11)},\quad\quad
\ket {{\bf{v}}_o}_n = \sqrt{\frac 38} (V_{0n}^{(12)}-V_{0n}^{(21)})\0
\ee
It is interesting to remark that the parameter $\e$ appears always divided by
$b$, so that one could just as well absorb $\e$ into $1/b$ and say that
the expansion is in the parameter $1/b$ for large $b$. However to avoid
confusion it is useful to keep the two parameters distinct.

Now, it is immediate to see that
\be
T' = \left(\matrix {-1 + {\cal O}(\frac {\e}b) & {\cal O}(\sqrt{\frac {\e}b})\cr
{\cal O}(\sqrt{\frac {\e}b})& T+ {\cal O}(\frac {\e}b)}\right)\label{T'approx}
\ee
This is correct provided we can prove that the use of (\ref{X'approx}) to
compute
$T'$ makes full sense,
that is all the terms of the expansion in powers of $\sqrt{\frac {\e}b}$
are well defined. One can actually see that a naive expansion leads
to infinite coefficients. This is a well--known problem, pointed out for the
first time in \cite{MT}, which requires a regularization. A nice
way to introduce a regulator is to switch on a constant background $B$ field.
We will not do it here, but we quote the result: in the presence of
a $B$ field the infinities disappear, and the expansion (\ref{T'approx})
makes full sense. From this we deduce in particular that
\be
T'_{nm} = T_{nm} + {\cal O}(\frac {\e}b)\label{T'=T+e}
\ee
This result is used in Section 6.

Let us consider now another extreme expansion, that is the limit $\a'\to
\infty$. In just the same way as above, we can introduce an alias, $t$
($t>>1$) instead of $\e$. So, in particular,
\be
&& V^{(rs)}_{mn}  \rightarrow  V_{mn}^{(rs)} \0  \\
&& V_{m0}^{(rs)}  \rightarrow  \sqrt{t} V_{m0}^{(rs)} \label{Vt} \\
&& V_{00}  \rightarrow  t V_{00} \0
\ee
In this case $X'$ to the lowest orders of approximation becomes
\be
X' = \left(\matrix{1 +\frac 23 \frac 1{V_{00}}\frac {b}{t}& -\frac 23
\sqrt{\frac {2b}t } \bra {{\bf{v}}_e}\cr
-\frac 23
\sqrt{\frac {2b}t } \ket {{\bf{v}}_e}&
X- \frac 43\frac 1{V_{00}}(1-\frac 1{V_{00}} \frac b{2t})( \ket {{\bf{v}}_e} \bra {{\bf{v}}_e}
-\ket {{\bf{v}}_o} \bra {{\bf{v}}_o})}\right)\label{X'0approx}
\ee
The lowest order in this expansion is known as the tensionless
limit \cite{giulio}.
Also here one must be careful about the use of this expansion in calculating
$T'$. From eq.(\ref{X'0approx}) one finds that
\be
T'_{00} = 1+ {\cal O}({\frac bt})\label{T'00approx}
\ee

\section{The $\a'\to 0$ limit of $\check S^{'(c)}_{nm}$ and $\check S^{'(c)}_{0n}$}

In this Appendix we discuss the limit of the unconventional lump matrix
elements $\check S^{'(c)}_{nm}$ and $\check S^{'(c)}_{0n}$ by means of the diagonal basis.
According to (\ref{btoinf}), we speak interchangeably
of the $b\to\infty$ limit and the $\eta\to \infty$ one. When applying the
results of this Appendix to section 6, we understand that $1/b$ is
replaced everywhere by $\e/b$ with finite $b$.

As a preliminary step let us prove that
\be
\lim_{b\to \infty} \left(V_0^{(k)}\right)^2=
\delta(k)\label{deltak}
\ee
A rather informal way to see this is as follows. Looking at (\ref{V0k}) it is
easy to realize that the limit always vanishes provided $k\neq 0$. Therefore
the support of the limiting distribution must be at $k=0$. We can therefore
expand all the functions involved in $k$ around $k=0$ and keep the leading terms.
Since $\Re F_c(k) \approx 1.386...$ around this point, we can disregard
$\Re F_c(k)$ compared to $b/4$ in the $b\to\infty$ limit. Therefore we easily
find
\be
\lim_{b\to\infty}(V_0^{(k)})^2= \lim_{b\to\infty}=
\frac {\bar b}{\pi} \frac 1{1+\bar b^2k^2} \0
\ee
where $\bar b=b/8$. Now defining $\bar\e = 1/{\bar b}$, the limit becomes
\be
\lim_{\bar \e\to 0} \frac 1{\pi} \frac {\bar \e}{k^2+\bar \e^2}=\delta(k)
\label{limdelta}
\ee
according to a well--known representation of the delta function. We can also
show that
\be
(V_0^{(k)})^2= \delta(k)+ {\cal O}(1/b)\0
\ee

From now on we suppose that, in the $\int dk$ integrals , we are allowed
to replace
the integrands with their $1/b$ expansions, and that the results we
obtain are valid at least in an asymptotic sense. This attitude is
always confirmed by numerical approximations.

\subsection{Limit of $\check S_{mn}^{'(c)}$}

Let us rewrite the generating function for $V_{m}^{(k)}$ as follows:
\begin{equation}
F^{(k)}(z)=A^{(k)}{\it f}^{(k)}(z)-
\frac{(1-\nu(k))V_{0}^{(k)}}{\sqrt{b}}B(k,z)
\label{eq1}
\end{equation}
where
\begin{equation}
A^{(k)}=V_{0}^{(k)}\sqrt{\frac{2}{b}}k\left(\Re F_{c}(k)-\frac{b}{4}\right)
\label{eq2}
\end{equation}
and
\be
B(k,z)\!&=&\!\frac{2}{1-\nu(k)}\left[\Re F_{c}(k)+\frac{\pi}{2\sqrt{3}}
\frac{\nu(k)-1}{\nu(k)+1} + \frac{2i}{k} + {\rm log}(iz)- 2i\it{f}^{(k)}(z)
\right.\label{eq3}\\
\!&+&\!\left.{\rm LerchPhi}(e^{-4i{\rm arctan}(z)},1,1+\frac{k}{4i})
e^{-4i{\rm arctan}(z)}e^{-k{\rm arctan}(z)}\right]\0
\ee
From (\ref{eq1}) we can derive a useful expression for $V_{m}^{(k)}$:
\begin{equation}
V_{m}^{(k)}=A^{(k)}\frac{\sqrt{m}}{2\pi i}\oint dz
\frac{{\it f}^{(k)}(z)}{z^{m+1}}-\frac{(1-\nu(k))V_{0}^{(k)}}{\sqrt{b}}
\frac{\sqrt{m}}{2\pi i}\oint dz \frac{B(k,z)}{z^{m+1}}
\label{eq4}
\end{equation}
Since $v_{m}^{(k)}=\frac{\sqrt{m}}{2\pi i}\oint dz
\frac{{\it f}^{(k)}(z)}{z^{m+1}}$ and
$S^{'(c)}_{mn}=\int_{-\infty}^{\infty} dk \ t(k) V_{m}^{(k)} V_{n}^{(-k)}$ we get:
\be
\check S_{mn}^{'(c)}\!&=&\!\int_{-\infty}^{\infty} dk \ t(k)\left[A^{(k)}
A^{(-k)}v_{m}^{(k)}v_{n}^{(-k)}
-A^{(k)} V_{0}^{(k)} v_{m}^{(k)}(1-\bar{\nu}(k))\tilde{B}_{n}(-k)
\frac{1}{\sqrt{b}}\right.\label{eq5}\\
\!&-&\!\left.A^{(-k)} V_{0}^{(k)} v_{n}^{(-k)}(1-\nu(k))\tilde{B}_{m}(k)
\frac{1}{\sqrt{b}}+ (V_{0}^{(k)})^{2}(1-\bar{\nu}(k))(1-\nu(k))
\tilde{B}_{m}(k)\tilde{B}_{n}(-k)\frac{1}{b}\right]\0
\ee
where
\be
\tilde{B}_{m}(k)=\frac{\sqrt{m}}{2 \pi i} \oint dz \frac{B(k,z)}{z^{m+1}}\0
\ee

Now we want to take the limit of (\ref{eq5}) when $b\rightarrow \infty$.
To this end we notice the following:
\be
\lim_{b \to \infty}A^{(k)}A^{(-k)}\!&=&\!\lim_{b \to \infty}(V_{0}^{(k)})^{2}
\left(\frac{-2k^{2}}{b}\right)\left(\Re F_{c}(k)-\frac{b}{4}\right)^{2}
\0\\
&&=\lim_{x \to -\infty}\left(\frac{-k^{2}}{N(k)}\right)
\frac{x^{2}}{4+k^{2}x^{2}}=
\left(\frac{-k^{2}}{N(k)}\right)\frac{1}{k^{2}}=-\frac{1}{N(k)}\0
\ee
where $x=\left(\Re F_{c}(k)-\frac{b}{4}\right)$. When $k$ is very large
$\Re F_{c}(k)$ tends to (slowly) diverge, but the factor $t(k)$ in
the integrand of (\ref{eq5}) concentrates the integral in the small $k$
region.

We also need:
\be
&&\lim_{b \to \infty} \frac{A^{(k)}V_{0}^{(k)}}{\sqrt{b}}=
\sqrt{2}k\delta (k)\left(\frac{\Re F_{c}(k)}{b}-\frac{1}{4}\right)\0\\
&&\lim_{b \to \infty} \frac{A^{(-k)}V_{0}^{(k)}}{\sqrt{b}}
=-\sqrt{2}k\delta (k)\left(\frac{\Re F_{c}(k)}{b}-\frac{1}{4}\right)\0
\ee

Finally using these limits
\be
\lim_{b \to \infty}\check S_{mn}^{'(c)}\!&=&\! -\int_{-\infty}^{\infty}
\frac{dk}{N(k)} \ t(k)v_{m}^{(k)}v_{n}^{(-k)}\0\\
&&+\lim_{b\to\infty}
\int_{-\infty}^{\infty} dk \ t(k)\delta (k)(1-\bar{\nu}(k))
(1-\nu(k))\tilde{B}_{m}(k)\tilde{B}_{n}(-k)\frac{1}{b}\0
\ee
while the other integrals vanish because they contain the factor
$k \delta (k)$. Here we have used the fact that $\nu(0)=\bar\nu(0)=-1$ and
$\tilde B_m(0)$ is finite, for a straightforward calculation gives
\be
\tilde{B}_{m}(0)=\frac{\sqrt{m}}{2 \pi i} \oint dz
\frac{{\rm log}(1+z^{2})}{z^{m+1}}= \left\{ \begin{array}{ll}
         0 & \mbox{for m odd};\\
        \frac{\sqrt{2m}}{2}(-1)^{\frac{m}{2}+1}(\frac{m}{2}+1)!
    & \mbox{for m even}.\end{array}\right.
\ee

So we are left with:
\begin{equation}
\lim_{b \to \infty}\check S_{mn}^{'(c)}=S_{mn}
\label{eq6}
\end{equation}
This is the sliver. The corrections are of order $\frac{1}{b}$.

\subsection{Limit of $\check S_{0m}^{'}$}

In the rest of this appendix we would like to justify eq.(\ref{lowosc}).
The limit of $\check S_{0m}^{'(c)}$ can be computed the same way as before. We have:
\begin{equation}
\lim_{b \to \infty}\check S_{0m}^{'(c)}=
\lim_{b \to \infty} \int_{-\infty}^{\infty} dk \ t(k) V_{0}^{(k)}V_{m}^{(-k)}
=\lim_{b \to \infty} \left(\int_{-\infty}^{\infty} dk \
t(k) V_{0}^{(k)}A^{(-k)}v_{m}^{(-k)}+\frac{2}{\sqrt{b}}\tilde{B}_{m}(0)\right)
\label{eq7}
\end{equation}
The last term in the RHS of course vanishes in the limit $b\to\infty$,
while the first limit diverges, but, recalling (\ref{lowosc}), what we are  really
need to know is the limit of $\frac{\check S_{0n}^{'}}{1+s^{'}}$.
Using the fact that $1+s^{'} \approx 4 \eta \ {\rm log} \ \eta$ when
$b \rightarrow \infty$ ($b \approx 4 {\rm log} \ \eta$) and
that we can write $\check S_{0m}^{'}=\check S_{0m}^{'(c)}+\check S_{0m}^{'(d)}$ (factorization
into continuous and discrete parts) we have:
\be
&&\check S_{0m}^{'(d)}\approx 2 \eta \sqrt{2{\rm log} \ \eta}\0\\
&&\check S_{0m}^{'(c)}\approx\int_{-\infty}^{\infty}dk
\ t(k) v_{m}^{(-k)}(-\sqrt{2}k)(V_{0}^{(k)})^{2}
\left(\frac{\Re F_{c}(k)}{4{\rm log} \ \eta} -
\frac{1}{4}\right)2\sqrt{{\rm log} \ \eta}\0
\ee
\
Using these we get:
\be
\frac{\check S_{0m}^{'(c)}}{1+s^{'}} \approx
\int_{-\infty}^{\infty} dk \ t(k)
 v_{m}^{(-k)}   (\sqrt{2}k )\delta(k)
 \left(\frac{\Re F_{c}(k)}{4{\rm log} \ \eta} - \frac{1}{4}\right)
   \ \frac{1}{2\eta}=0\0
\ee
and
\be
 \frac{\check S_{0m}^{'(d)}}{1+s^{'}}\approx \frac 1{\sqrt{2\,\log\,\eta}}\0
\ee
Hereby the conclusion (\ref{lowosc}) follows.


\end{document}